\documentclass[journal]{IEEEtran}
\usepackage{ifpdf}
\usepackage{bm}
\usepackage{cite}
\ifCLASSINFOpdf
\usepackage{graphicx}
\else
\fi
\usepackage{amsmath}
\usepackage{amsfonts}
\usepackage{array}
\usepackage{url}
\usepackage{booktabs}
\usepackage{arydshln}
\usepackage{algorithm}
\usepackage[noend]{algpseudocode}
\usepackage{amsfonts}
\usepackage{tabularx}
\usepackage{amssymb}
\usepackage{multirow}
\usepackage{threeparttable}
\usepackage{makecell}
\usepackage{rotating}
\usepackage{multicol}
\usepackage{booktabs}
\usepackage{xcolor}
\usepackage{array}
\usepackage{subfigure}
\usepackage{verbatim}
\usepackage{lineno,hyperref}
\usepackage{pifont}
\usepackage{amsmath,bm}
\definecolor{c6}{HTML}{EA0000}
\definecolor{c5}{HTML}{00A600}

\begin{document}

\title{AD-Mamba: Architecture-Complementary Knowledge Distillation for Efficient and Robust Audio Classification}

\author{Junyu Wang, Tianrui Wang, Meng Ge, Longbiao Wang, Xiaobao Wang, Jianwu Dang
\thanks{
This work was supported by the National Natural Science Foundation of China under Grants U23B2053 and 62176182. \textit{(Corresponding authors: Longbiao Wang)}.

Junyu Wang, Tianrui Wang, Xiaobao Wang are with the Tianjin Key Laboratory of Cognitive Computing and Application, College of Intelligence and Computing, Tianjin University, Tianjin 300350, China (e-mail: \{junyu\_wang21, wangtianrui, wangxiaobao\}@tju.edu.cn).

Meng, Ge is with Saw Swee Hock School of Public Health, National University of Singapore, Singapore (e-mail: gemeng@nus.edu.sg).

Longbiao Wang is with the Tianjin Key Laboratory of Cognitive Computing and Application, College of Intelligence and Computing, Tianjin University, Tianjin 300350, China, and also with the Huiyan Technology (Tianjin) Company, Ltd., Tianjin 300350, China (e-mail: longbiao\_wang@tju.edu.cn).

Jianwu Dang is with the Shenzhen Institute of Advanced Technology, Chinese Academy of Sciences, Guangdong, China (e-mail: jdang@jaist.ac.jp).

}}

\maketitle

\begin{abstract}
Recent advancements in large-scale audio classification have been primarily driven by Transformer-based architectures due to their strong capacity in modeling long-range dependencies. However, their inherent quadratic complexity limits scalability and efficiency, especially when processing long or sparsely informative audio streams. Mamba, a recently proposed state space model (SSM), offers a promising alternative by enabling efficient long-range modeling with linear sequence scaling. Nevertheless, existing Mamba-based audio classification methods still lag behind mainstream Transformer-based models, suggesting that effective training strategies for fully exploiting the audio modeling potential of SSMs remain underexplored.
To address this issue, we propose AD-Mamba, an architecture-complementary multi-stage knowledge distillation framework that transfers the local modeling strength of CNNs and the global contextual modeling capability of Transformers into a Mamba-based student model. Specifically, the CNN-guided stages encourage the student to adapt to local spectro-temporal structures, while the Transformer-guided stage further refines global semantic discrimination and contextual modeling. Through this local-to-global distillation process, AD-Mamba combines the efficiency of state space models with the complementary strengths of CNN and Transformer teachers. Experimental results on AudioSet show that AD-Mamba achieves 50.3\% mean average precision (mAP) with only 30M parameters, outperforming representative Transformer-based and Mamba-based audio classification baselines with lower computational cost. Further evaluations under long-form audio, sparse-information, and noise-interference conditions show that AD-Mamba consistently outperforms representative CNN and Transformer baselines. When transferred to FSD50K, ESC-50, and DCASE16, the proposed model also achieves superior performance, demonstrating strong generalization across audio classification tasks. These results indicate that architecture-complementary distillation is an effective strategy for building efficient and robust audio classification models based on state space architectures.
\end{abstract}

\IEEEpeerreviewmaketitle

\section{Introduction}

Audio classification, a foundational task in speech and audio processing, aims to assign semantic labels to acoustic signals, playing a pivotal role in applications ranging from environmental sound recognition to music tagging and speech emotion analysis. In practical intelligent audio systems, models are often expected to recognize acoustic events from long, noisy, and sparsely informative audio streams under limited computational budgets. In the era of deep learning, convolutional neural networks (CNNs) \cite{CNN} and Transformer-based \cite{transformer} architectures have emerged as the two dominant frameworks for this task \cite{panns, passt, audiomae}. CNNs capitalize on their spatial hierarchical design and translation-equivariant inductive biases to efficiently extract localized features, adeptly capturing fine-grained patterns in the acoustic domain \cite{psla, MN-KD}. Nevertheless, their capacity to model long-range temporal dependencies remains inherently constrained. To overcome this limitation, complex architectural designs or temporal pooling strategies are often employed. However, these approaches can undermine the model’s ability to preserve detailed temporal information.

Conversely, Transformers have emerged as the mainstream standard for sequence modeling, primarily due to their self-attention mechanism that enables the capture of global context across sequences \cite{HTS-AT, data2vec, beats}. Despite their success, Transformers suffer from two significant drawbacks: quadratic computational complexity with respect to sequence length and an over-reliance on explicit positional encodings. These limitations may lead to scalability and generalization issues, especially when dealing with variable-length audio data.

State space models (SSMs), most notably the Mamba architecture, have recently emerged as a viable alternative to Transformers \cite{SSM, mamba}. Mamba offers a linear scaling solution for sequence processing while retaining the ability to model long-range dependencies effectively. By leveraging state transitions to implicitly encode temporal dynamics, Mamba obviates the need for explicit positional encodings, a design choice that harmonizes well with the inherently sequential nature of audio signals. Several studies have explored the application of Mamba in audio classification \cite{AuM, SSAM, SSAMBA, audiomamba}, demonstrating its computational and memory efficiency advantages over Transformers, particularly in handling extended audio sequences. However, a significant performance disparity persists: although Mamba-based models excel in computational efficiency, they often lag behind established Transformer-based methods in classification accuracy on standard benchmarks \cite{EAT, ASDA, CED}. This suggests that effective training strategies for fully exploiting the audio modeling potential of SSMs remain underexplored, particularly when adapting Mamba to complex spectro-temporal classification tasks.

Recently, knowledge distillation has been successfully applied in the speech domain as an effective means of transferring knowledge from a larger, often pre-trained teacher model to a more compact student model \cite{distillspeech2, distillspeech3, distillspeech4}. The technique has also achieved remarkable success in other domains such as natural language processing (NLP) and computer vision (CV) \cite{transformertossm, distillCV1, distillCV2}. In the context of audio classification, traditional knowledge distillation methodologies predominantly rely on a single large-scale CNN or Transformer model to supervise the training of a more efficient student model. For instance, the recently proposed CED model \cite{CED} demonstrates that knowledge distillation can simultaneously improve performance and reduce model complexity, highlighting its potential to achieve excellent performance with lower computational demands.

Furthermore, several studies have explored the potential of cross-architectural knowledge distillation, demonstrating that transferring knowledge between different model architectures can yield substantial performance gains \cite{kd-cnn, transformertossm2}. For instance, bidirectional knowledge distillation between Transformer and CNN models has been shown to enable student models to outperform their original teacher counterparts \cite{CMKD}, allowing the student to inherit the unique advantages of each architecture: the local feature extraction of CNNs and the global contextual modeling of Transformers. Preliminary investigations into Mamba-based distillation have also shown promising results \cite{transformertossm, distillmamba}. In audio classification, DASS \cite{DASS} adopts a simple configuration where a Mamba-based student is distilled from a Transformer teacher, leading to clear performance improvements over both non-distilled model and even the teacher model itself. Moreover, its distilled audio SSMs have exhibited enhanced duration scalability, further underscoring the potential of Mamba in knowledge distillation. However, current research has yet to explore how to fully leverage the complementary strengths of different pre-trained models, specifically their respective expertise in local and global modeling, to effectively guide Mamba-based audio classifiers. In particular, it remains unclear how local structural knowledge from CNNs and global contextual knowledge from Transformers can be progressively integrated into an efficient SSM backbone for robust audio classification.

Building upon these foundations, we present AD-Mamba, a Mamba-based audio classification architecture trained via an architecture-complementary multi-stage knowledge distillation approach. This framework is designed to sequentially transfer the local inductive biases of CNNs and the global attention patterns of Transformers into the Mamba backbone, following a local-to-global knowledge transfer process. Specifically, the CNN-guided stages encourage the student to adapt to local spectro-temporal structures, whereas the Transformer-guided stage further refines global semantic discrimination and contextual modeling. In this way, AD-Mamba aims to combine the linear-time sequence modeling capability of Mamba with the complementary strengths of heterogeneous teacher architectures. Given the lack of standardized pre-training methods for audio SSMs and the proven efficacy of cross-modal pre-training in enhancing audio tagging \cite{vmamba, DASS}, we leverage ImageNet-pretrained SSM parameters to initialize weights of AD-Mamba, enabling better representation learning.

Extensive experiments on AudioSet demonstrate that AD-Mamba achieves 50.3\% mean average precision (mAP) using only 30M parameters, surpassing prior Transformer-based state-of-the-art (SOTA) models such as CED \cite{CED} (86M) and ASDA (93M) \cite{ASDA}. We also evaluate its capabilities in long-audio processing, sparse-information robustness, and noise robustness. Results show that our method outperforms CNN and Transformer baselines under these evaluation conditions, indicating that SSMs with multi-stage distillation exhibit stronger temporal adaptability and stable representation extraction. In downstream tasks, including sound event tagging (FSD50K) \cite{FSD50K}, environmental sound classification (ESC-50) \cite{ESC-50}, and acoustic scene classification (DCASE16) \cite{DCASE16}, AD-Mamba surpasses strong baseline models, including those trained at 32\,kHz, despite utilizing only a 16\,kHz sampling rate, highlighting its strong cross-task generalization capability.

Our core contributions can be summarized as follows:
\begin{itemize}

\item We present an architecture-complementary multi-stage knowledge distillation framework to progressively transfer the local feature extraction of CNNs and the global contextual modeling capabilities of Transformers into a Mamba-based student model, while preserving the inherent architectural efficiency of Mamba.

\item We design AD-Mamba as an efficient Mamba-based audio classification model that integrates local enhancement modules and multi-directional state space modeling to better adapt SSMs to two-dimensional audio spectrograms.

\item Experiments show that AD-Mamba outperforms CNN and Transformer baselines in long-audio processing, sparse-information robustness, and noise robustness, demonstrating its suitability for long and noisy audio classification scenarios.

\item AD-Mamba achieves state-of-the-art performance with 65\% fewer parameters than previous best models while requiring only 24GB GPU memory during training, making high quality audio classification more accessible.

\end{itemize}

Section \ref{related} reviews related work on audio classification architectures and knowledge distillation. Section \ref{preliminary} introduces the fundamental concepts of SSM. Section \ref{method} details the AD-Mamba framework, including its multi-stage distillation pipeline and architectural design. Section \ref{experiment_sec} presents experimental setups, benchmark datasets, comparative results, and ablation studies. Section \ref{downstream} demonstrates the performance of the model on various downstream tasks. Finally, Section \ref{conclusion} summarizes the contributions and outlines future research directions.

\section{Related work}
\label{related}

\subsection{Audio classification}
As a fundamental technology in auditory scene analysis, audio classification aims to assign semantic labels to acoustic signals, with wide-ranging applications in domains such as healthcare monitoring and industrial anomaly detection. Early deep learning approaches predominantly relied on convolutional neural networks (CNNs) or hybrid CNN-Attention architectures to model spectro-temporal patterns. A paradigm shift occurred with the Audio Spectrogram Transformer (AST) \cite{ast}, which demonstrated that pure attention-based models could outperform CNNs in this domain. This breakthrough inspired numerous supervised advancements: HTS-AT \cite{HTS-AT} replaced standard Vision Transformers (ViTs) \cite{vit} with hierarchical Swin Transformers \cite{swintrans} to enhance multi-scale modeling, while PaSST \cite{passt} optimized computational efficiency by reducing input sequence lengths through patch removal. The advent of self-supervised learning (SSL) in natural language processing (NLP), such as BERT \cite{BERT} and GPT \cite{gpt}, and computer vision (CV), such as BYOL \cite{BYOL} and MAE \cite{MAE}, has since catalyzed a new research frontier. AudioMAE \cite{audiomae} and MSM-MAE \cite{MSM-MAE} adapted masked autoencoding to audio data, achieving state-of-the-art (SOTA) classification performance that surpassed supervised baselines across multiple datasets. BEATs \cite{beats} extended this framework to discrete token representations akin to HuBERT \cite{hubert}, though these improvements were accompanied by substantial computational overhead. To address this, EAT \cite{EAT} introduced a multi-teacher BYOL variant with high masking ratios, enabling efficient training on as few as four RTX 3090 GPUs. Subsequent innovations, such as ASDA \cite{ASDA}, tackled the imbalance of attention distribution in ViT through differential attention mechanisms.

Despite these advancements, the quadratic computational complexity of Transformers remains a fundamental bottleneck for long-input scenarios. The recent Mamba architecture \cite{mamba}, a selective state space model (SSM), offers a compelling alternative with linear-time complexity. Gu et al. \cite{mamba2} further established a close connection between Mamba and attention mechanisms in Mamba2, validating its efficacy for long-sequence modeling. However, standard unidirectional SSM scans may be less effective on spatially structured data, prompting the development of directional variants: window-based \cite{windowmamba}, bidirectional \cite{visionmamba}, and quad-directional \cite{vmamba} SSMs. These designs process each scan direction through separate selective-scan blocks before fusion, preserving spatial consistency in the input. In speech processing, Mamba has achieved notable success in tasks like source separation \cite{dpmamba, spmamba, mambaasrandsp}, enhancement \cite{mambaseunet, semamba, mambase2}, and speech recognition \cite{ASRmamba, ASRmamba2, ASRmamba3, ASRmamba4}. For audio classification, pioneering works such as SSAMBA \cite{SSAMBA}, Audio Mamba \cite{SSAM}, AuM \cite{AuM}, and DASS \cite{DASS} have showcased the potential of SSMs. However, compared with strong Transformer-based audio classifiers, there is still a significant performance gap, especially in complex multi-label scenarios. This gap suggests that effective architectural adaptation and training strategies for Mamba-based audio classification remain insufficiently explored.

\subsection{Knowledge distillation}
Knowledge distillation is a training paradigm initially proposed by Hinton et al. \cite{distill}, which transfers knowledge from complex teacher models to lightweight student models. While it was originally developed to address model deployment challenges in resource-constrained scenarios, it has since evolved into a crucial optimization technique across various domains, including NLP \cite{transformertossm}, CV \cite{distillCV2}, and speech processing \cite{distillspeech}. In NLP, it has facilitated the transfer of capabilities from large-scale teacher models to more compact student models, such as DeepSeek-MoE \cite{deepseekmoe} and Qwen3-235B \cite{qwen}, while preserving much of the original model capability. Similarly, in CV, DeiT \cite{deit} pioneered the use of CNN teachers to train lightweight ViT students, achieving comparable accuracy with substantially fewer parameters.

For audio classification, knowledge distillation research has diverged into two primary directions: intra-architectural distillation and cross-architectural distillation. Intra-architectural approaches, such as CED by Dinkel et al. \cite{CED}, employ Transformer-to-Transformer distillation to transfer knowledge from strong teacher models into compact audio classifiers, achieving SOTA performance on the AudioSet dataset. This demonstrates that intra-architectural distillation can effectively streamline knowledge transfer within the same model family. Conversely, cross-architectural distillation studies \cite{MN-KD, CMKD} have explored the transfer of knowledge from high-performance Transformers, such as the Audio Spectrogram Transformer (AST), to lightweight CNNs, resulting in efficient models with reduced parameter counts yet comparable or superior predictive accuracy. Experiments in \cite{DASS, CMKD} further validate that cross-architectural knowledge distillation not only maintains performance but also enhances training stability and mitigates overfitting. 

Notably, research on knowledge distillation involving the Mamba architecture remains scarce. Existing efforts primarily focus on one-dimensional sequential data in the NLP domain. For example, studies \cite{transformertossm, transformertossm2} concentrate on transferring knowledge from LLMs based on Transformers to Mamba, achieving competitive performance with lower computational resources and demonstrating the feasibility of training efficient SSM students under the supervision of attention-based teacher models. However, when dealing with multi-dimensional data such as audio spectrograms or images, traditional SSMs face limitations in capturing spatial information, which has hindered the progress of Mamba-based knowledge distillation research. Although \cite{DASS, distillCV3} made preliminary attempts to assist Mamba-based student training through distillation, these efforts still leave room for improvement when compared with strong domain-specific baselines. Particularly in the field of audio classification, there is still limited evidence that Mamba-based models can match or surpass strong Transformer-based classifiers on large-scale multi-label benchmarks. In this paper, we introduce the AD-Mamba framework, which leverages architecture-complementary multi-stage distillation to integrate the complementary strengths of diverse network types and capitalize on the inherent advantages of SSMs in efficiently learning long-sequence features, thereby addressing this gap.

\section{Preliminary}
\label{preliminary}
State space models (SSMs) offer an efficient alternative to self-attention mechanisms for sequence modeling, leveraging linear-time recursive and convolutional representations. Inspired by classical continuous-time linear time-invariant (LTI) systems, the core formulation of a continuous-time SSM is defined by the following differential equations:
\begin{equation}
\begin{aligned}
h'(t) &= Ah(t) + Bx(t) \\
\end{aligned}
\end{equation}
\vspace{-1.45em}
\begin{equation}
\begin{aligned}
y(t) &= Ch(t) + Dx(t) \\
\end{aligned}
\end{equation}
where \(h(t) \in \mathbb{R}^{N}\) represents the hidden state, \(x(t) \in \mathbb{R}\) and \(y(t) \in \mathbb{R}\) denote the input and output signals, respectively. The system is parameterized by the state matrix \(A \in \mathbb{R}^{N \times N}\), input matrix \(B \in \mathbb{R}^{N \times 1}\), output matrix \(C \in \mathbb{R}^{1 \times N}\), and skip connection \(D \in \mathbb{R}\).

To handle digital discrete signals common in modern computer systems, continuous SSMs are discretized using techniques like zero-order hold (ZOH), which assumes constant input values between sampling instants. For a sampling interval \(\Delta t\), the ZOH discretization yields: 
\begin{equation}
\begin{aligned}
h_{k+1} &= A_dh_k + B_dx_k \\
\end{aligned}
\end{equation}
\vspace{-1.45em}
\begin{equation}
\begin{aligned}
y_k &= C_dh_k + D_dx_k \\
\end{aligned}
\end{equation}
where \(A_d = e^{(\Delta t A)}\), \(B_d = (\int_{0}^{\Delta t} e^{As} ds) \cdot B\), \(C_d = C\), and \(D_d = D\). This recurrent formulation enables SSMs to process discrete sequences with linear complexity in sequence length.

Traditional time-invariant SSMs (e.g., S4 \cite{s4}) face limitations in modeling non-stationary dynamics (e.g., time-varying audio signals), as their fixed system matrices fail to adapt to evolving input characteristics. Mamba (S6) addresses this by introducing a time-varying selective scan mechanism, where the input coupling matrix $\mathbf{B}$ and output projection matrix $\mathbf{C}$ are parameterized as functions of the input $\mathbf{x}_k$, while the state transition matrix $\mathbf{A}$ remains time-invariant (e.g., a diagonal matrix with decay factors) to ensure stable long-range dependency modeling.  

The Mamba SSM computes outputs via a causal selective scan, where the contribution of each input $\mathbf{x}_i$ (for $i \leq k$) to the current output $y_k$ is weighted by a product of state transition factors from $i$ to $k$:  
\begin{equation}
y_k = \mathbf{C}_k \underbrace{\left( \sum_{i=1}^k \left( \prod_{j=i+1}^k \bar{A}_j \right) \bar{B}_i \mathbf{x}_i \right)}_{\text{context-aware memory}} + \mathbf{D} \mathbf{x}_k
\end{equation}  
Here, $\bar{A}_j$ represents a discretized state transition factor that controls the degree of information retention, $\prod_{j=i+1}^k \bar{A}_j$ captures the decay of information from time step $i$ to $k$, and $\bar{B}_i$ adapts to the input $\mathbf{x}_i$ to emphasize relevant features.  

This formulation enables Mamba to maintain $O(L)$ time complexity, while dynamically adjusting to non-stationary patterns, making it well-suited for long-duration audio signals with time-varying spectro-temporal characteristics.  

\begin{figure}[t]
  \centering
  \vspace{0cm}
  \includegraphics[width=0.8\linewidth]{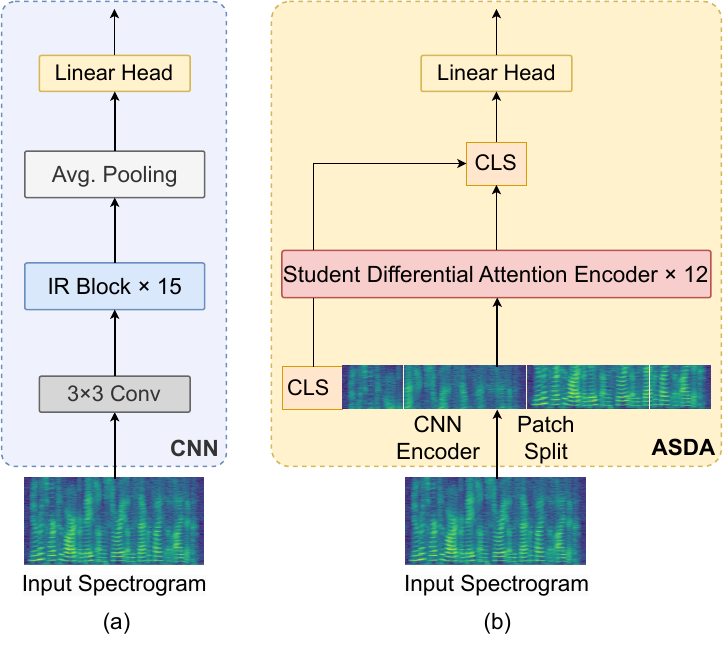}
  \vspace{0cm}
  \caption{Illustration of the complementary teacher architectures used in our distillation framework. (a) The CNN-based teacher model adopts a dynamic CNN backbone composed of 15 inverted residual (IR) blocks, providing local spectro-temporal modeling bias. (b) The Transformer-based teacher leverages the ASDA (Audio Spectrogram Differential Attention) architecture, which incorporates a patch-split input pipeline and a differential attention encoder for capturing global dependencies in audio spectrograms.
}
  \label{figure1}
  \vspace{-0.3cm}
\end{figure}

\section{Architecture-complementary multi-stage distillation for AD-Mamba}
\label{method}
This section presents the architecture-complementary multi-stage knowledge distillation framework and the proposed AD-Mamba model architecture. Specifically, Section~\ref{conv and trans teacher} introduces the CNN- and Transformer-based teacher models, focusing on their architectural designs and complementary feature modeling capabilities; Section~\ref{architecture} describes the proposed AD-Mamba student model, detailing its selective state space mechanisms and adaptive designs tailored for audio classifications; and Section~\ref{knowledge distillation} elaborates on the distillation pipeline, including the procedures and configurations of the local-to-global multi-stage training strategy.

\begin{figure*}[t]
\centering
\vspace{-0.1cm}
\includegraphics[width=0.8\linewidth]{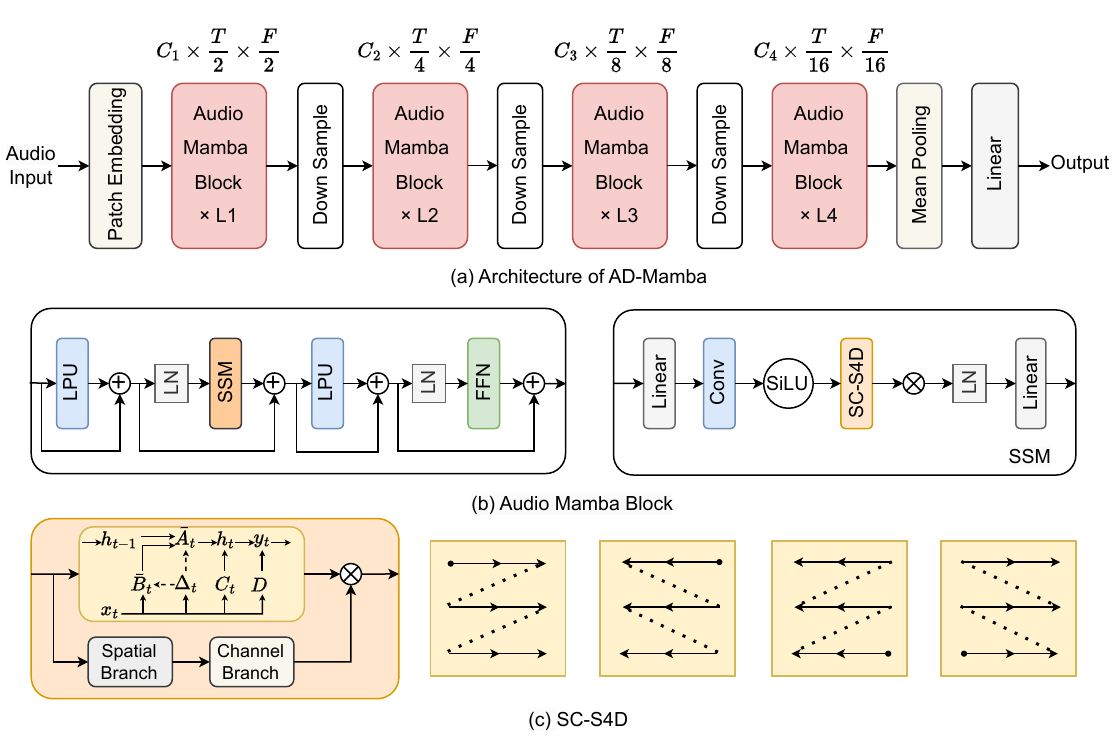}
\vspace{0cm}
\caption{Overview of the proposed AD-Mamba architecture for audio spectrogram classification. The model progressively extracts multi-scale spectro-temporal representations through patch embedding, stacked Audio Mamba Blocks, and intermediate downsampling operations. Each Audio Mamba Block integrates state space modeling, feed-forward transformation, and local perception units (LPUs) to combine long-range sequence modeling with local spectro-temporal perception}
\vspace{-0.1cm}
\label{figure2}
\end{figure*}

\subsection{CNN and Transformer teacher models}
\label{conv and trans teacher}

We adopt the dynamic CNN architecture proposed in \cite{kd-cnn} as the teacher model, as shown in Figure~\ref{figure1}(a), which integrates dynamic convolution, dynamic activation functions, and coordinate attention into inverted residual (IR) blocks, and enables efficient global context modeling via a cross-block parameter sharing mechanism. In our framework, this CNN teacher mainly provides local spectro-temporal modeling knowledge for the Mamba-based student. To adapt to this study, we adjust the audio input processing pipeline from 32\,kHz (original) to 16\,kHz based on the open-source code.
The model is built upon the MobileNetV3 \cite{mobilenetv3} pretraining framework. For an input audio signal of duration $t$ seconds, it first converts the signal into 128-dimensional log-mel spectrogram features, yielding a spectral input of size $128 \times 100t$. Subsequently, a 3×3 convolution layer (stride 2×2, 16 output channels) reduces the dimension to $64 \times 50t \times 16$ for initial feature extraction. Then, 15 IR blocks are sequentially applied to leverage the local receptive field and translation invariance of CNNs for learning multi-scale audio features. Finally, after global average pooling, two fully connected layers are appended, with softmax (for single-label scenarios) or sigmoid (for multi-label scenarios) activation functions applied respectively to generate predicted classification labels.

For the Transformer-based teacher model, we employ the recently proposed ASDA (Audio Spectrogram Differential Attention) architecture \cite{ASDA}, as illustrated in Figure~\ref{figure1}(b). This teacher is used to provide global contextual and semantic discrimination knowledge for the student model. Designed for self-supervised representation learning, ASDA employs a multi-student single-teacher framework during pre-training to mitigate feature instability and computational burden. Specifically, the student model applies a block-wise random masking strategy, where only unmasked patches are fed into the encoder. A CNN decoder then reconstructs the masked segments using the encoder outputs. In contrast, the teacher model processes fully unmasked patch embeddings and updates via an exponential moving average (EMA) strategy. The core innovation of ASDA is its differential attention mechanism, which reformulates the standard self-attention computation through dual-softmax operations, effectively alleviating inefficient attention allocation issues prevalent in standard ViTs.

The input processing pipeline begins by converting a $t$ seconds audio signal into a 128-dimensional log filter bank spectrogram ($128 \times 100t$) using Hanning-windowed Fourier transforms with 10ms frame shifts. This spectrogram is then partitioned into non-overlapping patches, each linearly projected into a 768-dimensional embedding space. To compensate for the Transformer's lack of inherent positional sensitivity, we augment these embeddings with fixed 1D positional encodings and prepend a learnable CLS token to capture global utterance-level characteristics. These processed embeddings are then fed into a 12-layer Transformer encoder with the differential attention mechanism. During fine-tuning, only the encoder is retained, with a linear classification layer (sized by the number of downstream classes) added to generate predicted labels.

\subsection{AD-Mamba}
\label{architecture}

The detailed framework of the proposed AD-Mamba model is illustrated in Figure \ref{figure2}. For an input audio signal, we first transform it into a 2D log-Mel spectrogram representation \(X \in \mathbb{R}^{T \times F}\) following the same preprocessing pipeline used for the teacher models. The spectrogram is then processed through a patch embedding layer composed of 2D convolutions, generating initial feature maps with dimensions \(C_1 \times \frac{T}{2} \times \frac{F}{2}\). Next, multiple Audio Mamba blocks are employed to model spectro-temporal representations in this spatial dimension. To enable the model to perceive information at different scales, we also perform three down-sampling operations, resulting in feature maps with dimensions of \(C_2 \times \frac{T}{4} \times \frac{F}{4}\), \(C_3 \times \frac{T}{8} \times \frac{F}{8}\), and \(C_4 \times \frac{T}{16} \times \frac{F}{16}\). Finally, the output of the last Audio Mamba block undergoes average pooling and is then passed to a linear classification layer to obtain the model-predicted labels.

\begin{figure}[t]
  \centering
  \includegraphics[width=0.9\linewidth]{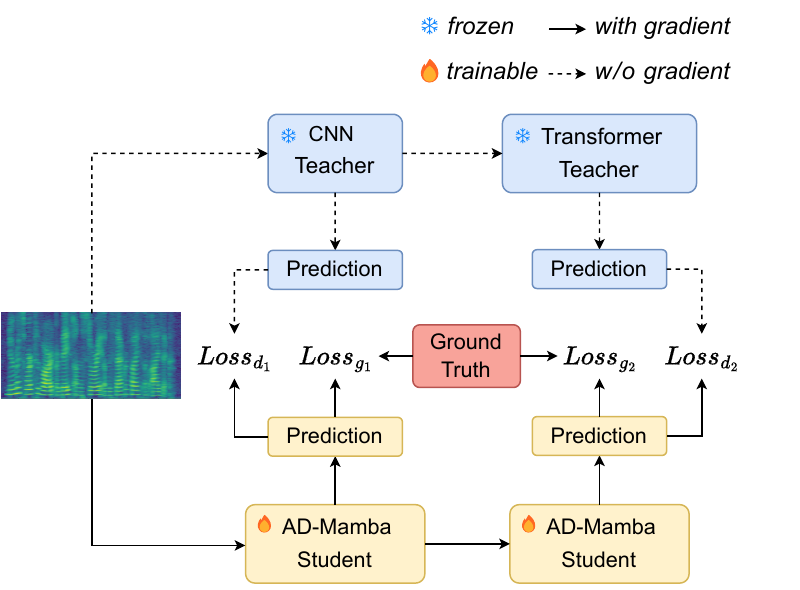}
  \vspace{0.2cm}
  \caption{Overall framework of the proposed architecture-complementary multi-stage knowledge distillation process. The CNN-guided stages first transfer local spectro-temporal modeling knowledge to the Mamba-based student, while the Transformer-guided stage further refines global semantic discrimination and contextual modeling. The student model is continuously optimized with both supervised classification loss and distillation loss throughout the sequential training process.}
  \label{figure3}
  \vspace{-0.3cm}
\end{figure}

The structure of the Audio Mamba Block is shown in Figure \ref{figure2}(b), which mainly consists of two local perception units (LPUs) \cite{LPU} for extracting local spectrogram patterns, a modified state space model (SSM) for capturing long-range dependencies, and a feed-forward neural network (FFN) for enhancing channel-wise feature transformation. Unlike standard SSM implementations, we remove the linear gating mechanism to reduce model complexity while maintaining training stability. Although the scanning mechanism in S6 can be well applied to one-dimensional temporal data, it may be insufficient for modeling two-dimensional time-frequency structures when processing audio spectrograms. To address this drawback, we adopt a quad-directional selective scanning approach inspired by image processing techniques \cite{vmamba}, as illustrated in Figure \ref{figure2}(c), so that S6 can better adapt to spectrogram data while preserving its computational efficiency. To further enhance the ability of selective scanning to capture meaningful patterns, we also introduce two specialized branches that work in conjunction with the selective scanning mechanism. The spatial branch employs pooling operations and convolutional layers to identify important regions in the time-frequency domain, while the channel branch focuses on learning inter-channel relationships through squeeze-and-excitation operations. Both branches utilize sigmoid activation functions to generate adaptive reweighting coefficients that are multiplied with the SSM outputs, allowing the model to dynamically emphasize the most relevant features with minimal additional parameters. In this way, the mechanism can allocate larger responses to more significant parts with only a minimal increase in parameters, optimizing the extraction of valuable features from the spectrogram.

\subsection{Knowledge distillation}
\label{knowledge distillation}

In this work, we aim to leverage pre-trained CNN- and Transformer-based audio classification models, trained with extensive computational resources, to facilitate the training and performance enhancement of a Mamba-based audio classification model via knowledge distillation. Unlike conventional knowledge distillation in audio domains, which typically employs a fixed teacher throughout training, our method adopts a progressive, multi-teacher distillation strategy tailored for state space models. Prior approaches, such as the framework introduced in \cite{kd-cnn}, leverage a Transformer as the sole teacher model to supervise the training of a CNN-based student. This design facilitates the transfer of global, data-driven representational capacity from the Transformer to the CNN, while preserving the inherent strengths of CNN models in local feature extraction and computational efficiency. However, such a single-teacher paradigm limits the scope of representational diversity that can be inherited by the student model.

In contrast, our framework is built upon the Mamba architecture, a state space model known for its inherent capability to model long-range dependencies and implicitly capture sequential order without relying on explicit positional encodings. We propose to sequentially distill knowledge from both CNN and Transformer teachers into Mamba, as shown in Figure \ref{figure3}. Specifically, the CNN teacher is first employed to guide the Mamba student toward local spectro-temporal structure modeling, capturing the spatially localized inductive biases that CNNs excel at. Subsequently, a Transformer teacher is used to enrich Mamba’s ability to model global contextual information through attention-based patterns. This staged distillation approach allows Mamba to inherit the complementary strengths of both teacher models while maintaining its native efficiency and scalability. 

During training, the teacher models are fixed in evaluation mode to prevent parameter updates. Both the teacher and student networks receive identically augmented audio spectrograms to ensure consistent input semantics. The student model learns to mimic the teacher outputs by aligning its logits with those of the teacher, thereby acquiring the latter’s discriminative structures and processing patterns. The overall loss function for student training is a weighted combination of the ground-truth classification loss and distillation loss:

\begin{equation}
\mathcal{L} = (1-\lambda)Loss_g(O_s, L) + \lambda Loss_d\left(O_s, \frac{O_t}{\tau}\right)
\end{equation}
where $O_s$ and $O_t$ denote the logits from the student and teacher models, respectively; $L$ denotes the ground-truth labels; and $\tau$ is a temperature parameter used to soften the teacher logits. The classification loss $Loss_g$ is implemented using binary cross-entropy (BCE) for multi-label classification, while the distillation loss $Loss_d$ is computed as the Kullback-Leibler (KL) divergence between the student and softened teacher outputs. The balancing coefficient $\lambda \in [0,1]$ controls the relative contribution of the two loss terms.

\section{Pre-training experiments and results}
\label{experiment_sec}

In this section, we evaluate the performance of the proposed method on large-scale audio event classification tasks, which serve as a critical benchmark for audio classification models and provide an important foundation for downstream audio captioning and understanding tasks. We leverage the AudioSet dataset, a widely used large-scale benchmark for audio classification, to train our models. Two model variants with different parameter configurations are employed: AD-Mamba(S), and AD-Mamba(M).

\subsection{Datasets}

AudioSet is one of the largest publicly available multi-label dataset for audio classification, consisting of approximately 2 million 10-second audio clips. Each clip is annotated with one or more labels spanning a taxonomy of 527 sound event classes, covering a wide range of acoustic scenes and source types. As AudioSet is collected from YouTube, some audio clips have become invalid over time. To ensure fair comparisons with existing methods, we utilize the publicly accessible version, which includes 1,912,134 available samples. For evaluation, we adopt the official evaluation set, which consists of 18,884 audio clips and shares the same 527 label classes. Given the multi-label nature of the dataset, we adopt mean average precision (mAP) as the evaluation metric.

\subsection{Pre-training experimental setup}

\subsubsection{Data processing}
\label{data processing}

To ensure fair comparison with prior baselines \cite{atst, ASDA}, all audio inputs are resampled to 16\,kHz. Log-mel spectrograms are extracted using a 25 ms window length and 10 ms hop size. We adopt standard data augmentation strategies during pretraining, including SpecAugment, random noise injection, mixup, and time-axis rolling \cite{psla}. Specifically, frequency and time masking lengths are set to 48 and 192, respectively, and the mixup coefficient is set to 0.5. These augmentations are applied consistently across all pretraining stages to enhance model performance and robustness.

\subsubsection{Training procedure}

We initialize the Mamba-based student models with ImageNet-pretrained state space model (SSM) parameters \cite{vmamba} rather than random initialization, a practice shown to be effective in previous works\cite{ast, DASS}. The training pipeline consists of three progressive stages:

Stage 1: The CNN teacher model is used to guide the core Mamba-based student backbone. To first adapt the representation capacity of the SSM backbone to the audio classification domain, the student model excludes the LPU, spatial branch, and channel branch, allowing this stage to focus on backbone-level local structural adaptation. This stage is trained for 100 epochs, each containing 100,000 audio clips.

Stage 2: The pretrained SSM backbone is frozen, and the LPU, spatial, and channel branches are added to the student model. The CNN teacher is retained to continue distillation, so that the newly introduced local enhancement modules can further capture fine-grained audio details under CNN-guided supervision. This stage leverages CNN inductive biases such as locality and translation equivariance, and also runs for 100 epochs with 100,000 samples per epoch.

Stage 3: All student model parameters are unfrozen, and the teacher is switched to the ASDA Transformer-based model to refine global semantic discrimination and contextual modeling. The self-attention mechanism in Transformers enables the integration of long-range dependencies and contextual information, complementing the local structural knowledge acquired in the previous stages. This final stage spans 200 epochs, also with 100,000 samples per epoch.

In all stages, training is conducted using the Adam optimizer \cite{adam} with a batch size of 48 and a maximum learning rate of 2e-4. We apply an exponential warm-up strategy for the first 5 epochs. Learning rates are decayed by a factor of 0.5 every 10 epochs in Stages 1 and 2, and every 20 epochs in Stage 3. The distillation temperature $\tau$ is set to 2.0 to soften the teacher’s output distribution, which facilitates more informative gradient signals compared to $\tau = 1.0$ by encouraging the student to learn the relative confidence between classes rather than mimicking hard labels. The distillation loss weight $\lambda$ is set to 0.5 to balance the distillation loss and classification loss.

\subsubsection{Model configuration}

We experiment with two model scales: AD-Mamba(S) and AD-Mamba(M). The primary difference lies in the number of AudioMamba blocks in each stage. For AD-Mamba(S), the number of blocks in layers L1–L4 is set to {2, 2, 8, 2}, while for AD-Mamba(M), it is {2, 2, 20, 2}. Both variants share the same channel configuration, with initial channel dimension $C_1=96$, and each downsampling stage doubles the number of channels.

\begin{table}[t]
    \centering
    \caption{Performance comparison with representative baseline models on AS-2M. KD indicates whether knowledge distillation is used. Floating point operations (FLOPs) representing the computational complexity, are calculated based on the model processing 10-second audio clips, with the results averaged over 100 repeated computations.}
    \footnotesize
    \setlength{\tabcolsep}{2.5pt}
    \begin{tabular}{c|ccccc}
        \hline
        Model  & Params & FLOPs & KD & Type & mAP  \\
        \hline
        PANN\cite{panns} & 81M & - & False & CNN & 43.9   \\
        PSLA\cite{psla} & 14M & - & False & CNN & 44.4  \\
        ConvNeXt\cite{ConvNeXt} & 28M & - & False & CNN & 47.1  \\
        MN\cite{MN-KD} & 68M & - & True & CNN & 48.3  \\
        DyMN\cite{kd-cnn} & 40M & - & True & CNN & 49.0  \\
        AST\cite{ast} & 86M & 130.6G & False & Transformer & 45.9  \\
        PaSST\cite{passt} & 86M & - & False & Transformer & 47.1  \\
        HTS-AT\cite{HTS-AT} & 31M & - & False & Transformer & 47.1  \\
        Audio-MAE\cite{audiomae} & 86M & - & False & Transformer & 47.3  \\
        EAT\cite{EAT} & 88M & 48.6G & False & Transformer & 48.6  \\
        ASDA\cite{ASDA} & 93M & 54.7G & False & Transformer & 49.0  \\
        BEATs\cite{beats} & 90M & - & True & Transformer & 48.6  \\
        ATST-C2F\cite{atst} & 86M & 48.5G & True & Transformer & 49.7  \\
        CED\cite{CED} & 86M & 45.3G & True & Transformer & 50.0  \\
        AuM\cite{AuM} & - & - & False & Mamba & 39.7  \\
        Audio Mamba\cite{audiomamba} & 40M & - & False & Mamba & 44.0  \\
        DASS\cite{DASS} & 49M & 22.2G & True & Mamba & 47.6  \\
        \hline
        AD-Mamba(S) & 30M & 12.7G & True & Mamba & 50.3  \\
        AD-Mamba(M) & 49M & 22.3G & True & Mamba & \textbf{50.6}  \\
        \hline
    \end{tabular}
    \vspace{-0.1cm}
    \label{as2m baseline comparison}
\end{table}

\subsection{Performance comparison with baselines}

\subsubsection{Performance comparison on AudioSet}

Table~\ref{as2m baseline comparison} presents the performance comparison between our proposed AD-Mamba framework and representative strong baseline models on the AS-2M dataset. Specifically, our AD-Mamba(M) model achieves 50.6\% mAP using only 49M parameters and 22.3G FLOPs, surpassing all previous methods, including the current best-performing CED model (50.0\% mAP with 86M parameters and 45.3G FLOPs), while using 43\% fewer parameters and 51\% less computational cost. Even more notably, the smaller variant AD-Mamba(S), with just 30M parameters and 12.7G FLOPs, still achieves 50.3\% mAP, outperforming all previous methods in performance in both accuracy and efficiency.

In comparison with prior Mamba-based models, the advantage becomes even more pronounced. Audio Mamba without distillation achieves only 44.0\% mAP, while DASS with distillation reaches 47.6\% mAP. Despite employing ImageNet-initialized parameters and knowledge distillation, these models still fall notably short of Transformer-based performance, and in some cases, even underperform modern CNN-based models such as MN (48.3\% mAP with 68M parameters) and DyMN (49.0\% mAP with 40M parameters).

Our proposed AD-Mamba addresses this gap by introducing an architecture-complementary multi-stage knowledge distillation framework, which sequentially transfers inductive biases and representational strengths from CNNs and Transformers into the Mamba student model. This process integrates local feature extraction capabilities and global context modeling capacity into the student, while leveraging Mamba’s inherent advantages in linear-time sequence processing and implicit positional modeling. The results suggest that Mamba, when guided through structured distillation, can match or exceed representative Transformer-based audio classifiers with substantially reduced computational cost, supporting the effectiveness of the proposed architecture-complementary design.

\begin{figure}[t]
  \centering
  \includegraphics[width=0.7\linewidth]{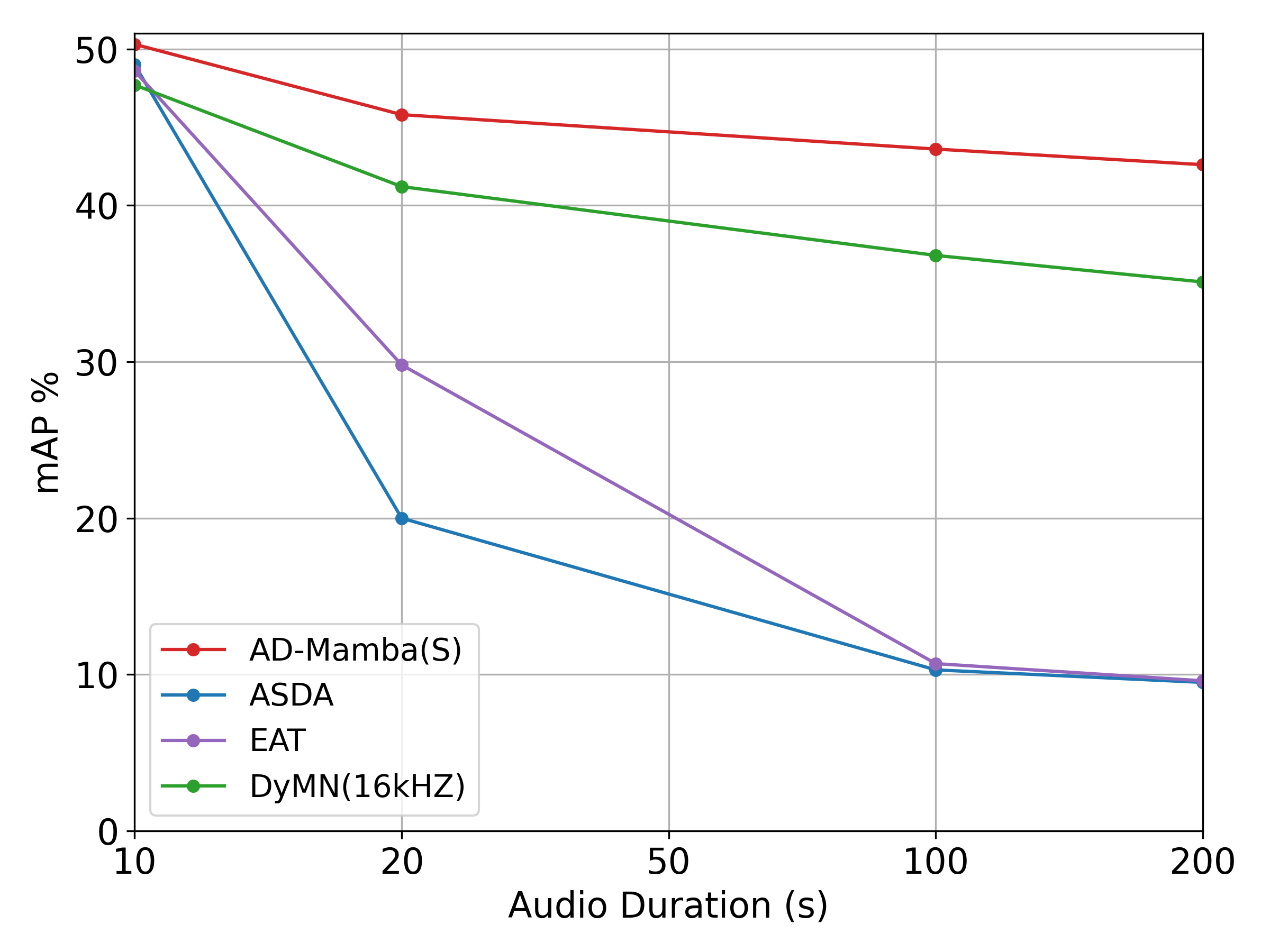}
  \vspace{-0.1cm}
  \caption{mAP performance comparison of audio classification models across varying audio durations.}
  \label{figure4}
  \vspace{-0.3cm}
\end{figure}

\subsubsection{Performance comparison on long-form audio classification}

Given the absence of a standardized benchmark for long-form audio classification in the community, we construct a controlled synthetic evaluation setting by concatenating multiple semantically related 10-second clips from the AudioSet evaluation set to generate longer audio inputs. We evaluate on audio durations ranging from 10 to 200 seconds, while ensuring that the concatenated clips share consistent or related sound labels. Figure~\ref{figure4} illustrates the mAP performance of our small model AD-Mamba(S) compared to three strong baselines under 16\,kHz audio: ASDA and EAT, two recent open-source Transformer-based models, and DyMN, a strong CNN model. Note that we exclude CED due to its reliance on 32\,kHz inputs, making fair comparisons infeasible. To ensure consistency, we re-implemented DyMN at 16\,kHz and used it as our CNN teacher model. The reproduced model achieves 47.7\% mAP on AudioSet, which provides a competitive CNN baseline under the same 16\,kHz setting.

The results demonstrate that AD-Mamba(S) consistently outperforms all baselines across varying audio durations, with the performance gap widening as input length increases. Transformer-based models, such as ASDA and EAT, suffer from scalability limitations due to their reliance on explicit positional encodings, which hinder generalization when encountering sequences significantly longer than those seen during training. At 10 seconds, AD-Mamba(S) achieves 50.3\% mAP, surpassing ASDA (49.0\%), EAT (48.6\%), and DyMN (16\,kHz) (47.7\%). Beyond 20 seconds, baseline models, especially ASDA and EAT, exhibit sharp performance degradation. This trend becomes more pronounced at 50 seconds and beyond, where AD-Mamba(S) maintains a competitive mAP of 43.6\%, while ASDA and EAT drop to near 10\%, and DyMN declines to 36.8\%. When the input length increases to 200 seconds, AD-Mamba(S) still sustains a strong mAP of 42.2\%, further widening the gap over ASDA (9.3\%), EAT (9.1\%), and DyMN (34.5\%). These findings highlight the better long-form stability of AD-Mamba(S) in extended audio classification scenarios.

\subsubsection{Robustness evaluation under zero and noisy padding}

To further evaluate the robustness of models against irrelevant or noisy content, we design a controlled experiment by randomly padding either zero-valued or noise segments at arbitrary temporal positions within 10-second effective audio clips. This setting aims to simulate scenarios where long audio streams contain only sparse informative segments, interleaved with irrelevant or noisy content, which is common in real-world deployments such as continuous monitoring and surveillance. Unlike standard long audio evaluation, this setup assesses both the temporal generalization ability of the model and its robustness under temporal sparsity and interference conditions.

\begin{figure}[t]
    \centering
    \includegraphics[width=0.7\linewidth]{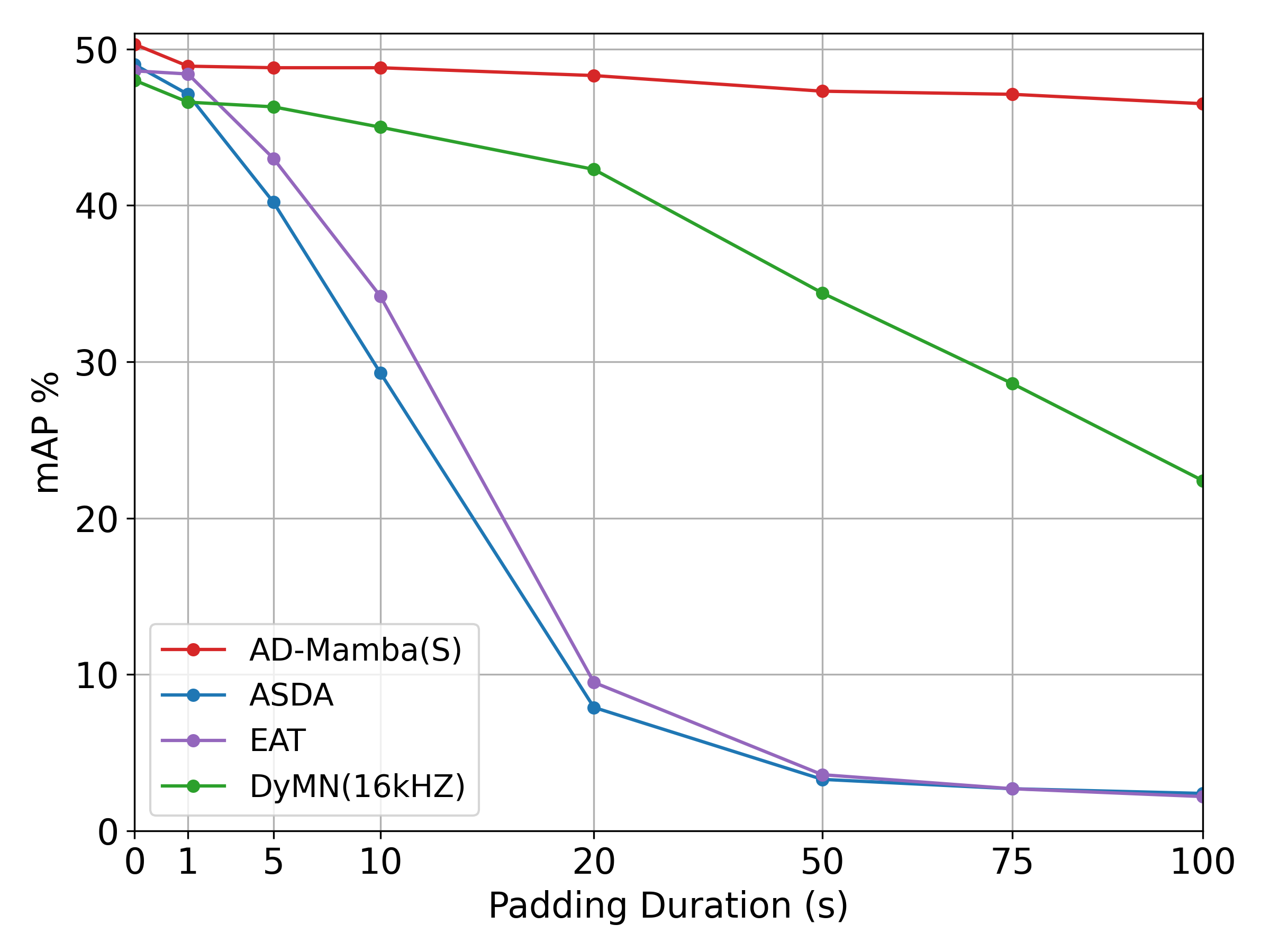}
    \vspace{-0.1cm}
    \caption{mAP performance comparison under increasing zero padding.}
    \label{figure5}
    \vspace{-0.3cm}
\end{figure}

We first investigate the zero-padding condition, where silent segments are inserted without altering the original content. As shown in Figure ~\ref{figure5}, Transformer-based models suffer from rapid performance degradation as the padding length increases. When only 5 seconds of silence are inserted, both ASDA and EAT underperform the CNN-based DyMN (16\,kHz). As the padding increases to 50 seconds, the performance of Transformer-based models nearly collapse (mAP $\approx$ 3.0\%), while DyMN (16\,kHz) drops to 34.4\%. In contrast, our AD-Mamba(S) retains a robust mAP of 46.5\% even when padding for 100 seconds, showing stronger stability under high-sparsity settings.

\begin{figure}[t]
    \centering
    \includegraphics[width=0.7\linewidth]{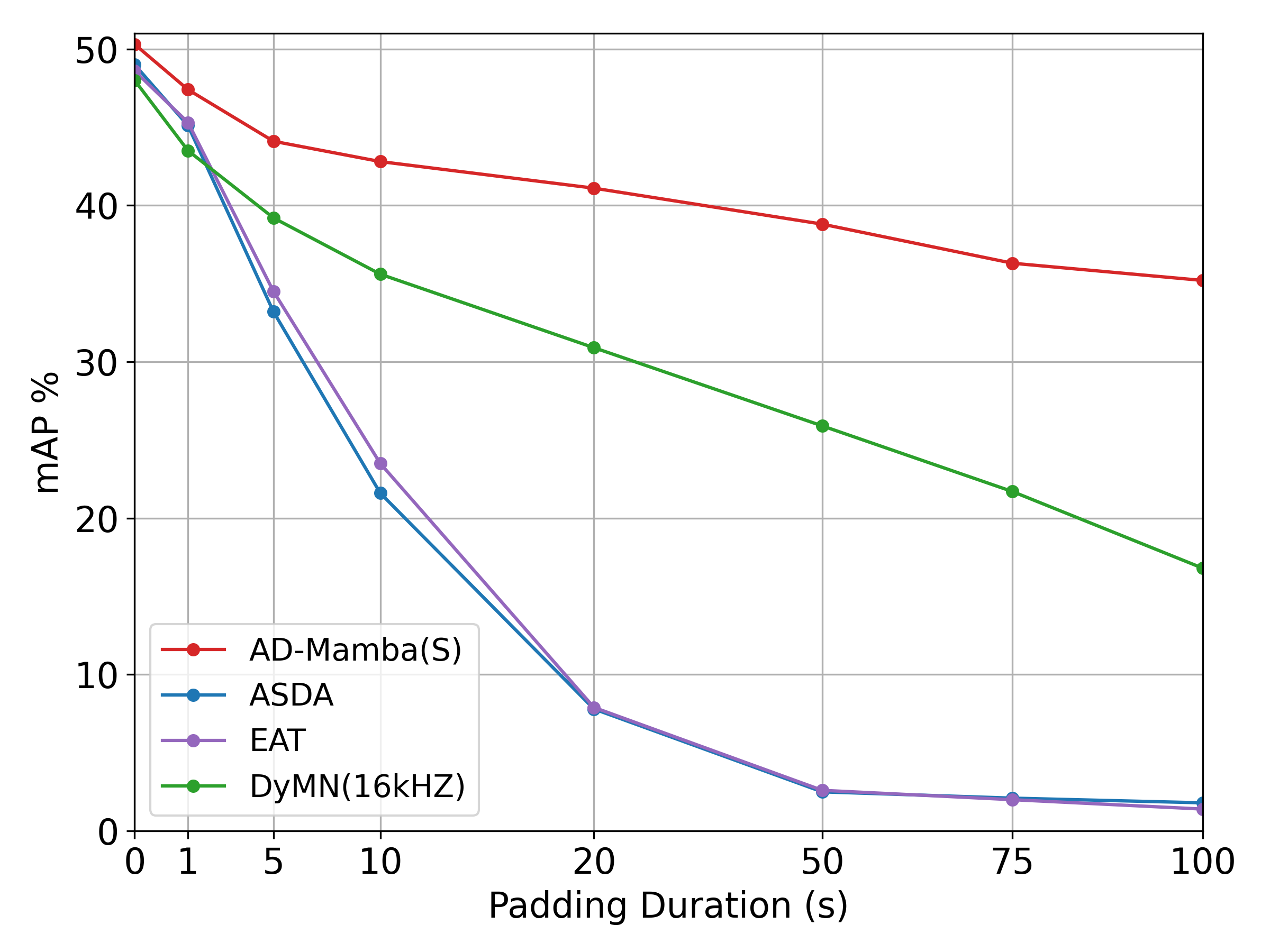}
    \vspace{-0.1cm}
    \caption{mAP performance comparison under increasing white noise padding.}
    \label{figure6}
    \vspace{-0.3cm}
\end{figure}

We then conduct a more challenging setting by injecting white noise with a 10 dB signal-to-noise ratio (SNR) at random positions, aiming to simulate common acoustic scenarios such as overlapping environmental sounds or unpredictable background interference. As illustrated in Figure~\ref{figure6}, the performance trends are similar to the zero-padding setting. With only 5 seconds of added noise, Transformer models degrade below CNN performance. When the duration of injected noise increases to 20 seconds, both ASDA and EAT degrade to below 10.0\% mAP, while DyMN (16\,kHz) drops to 30.9\%. In contrast, even with 100 seconds of noise padding, AD-Mamba(S) still maintains a strong mAP of 35.2\%, indicating stronger resistance to temporal noise interference.

These findings underscore the robustness of AD-Mamba(S), which benefits from the architecture-complementary multi-stage distillation strategy and the sequential modeling capability of Mamba. Compared to conventional Transformer and CNN designs, our model exhibits stronger noise resilience for long-duration audio inputs with interfering segments.

\subsection{Ablation study}
\subsubsection{Importance of ImageNet initialization and multi-stage knowledge distillation}

To analyze the contributions of ImageNet initialization (IN) and multi-stage knowledge distillation, we conduct ablation experiments as summarized in Table~\ref{ablation IN and KD}.

We begin by examining the effect of ImageNet initialization. Comparing 0-index (random initialization) with 6-index (ImageNet initialization), we observe a performance gain from 48.9\% to 50.3\% in mAP. This result highlights the benefit of leveraging large-scale visual pretraining to enhance the representational capacity of the Mamba backbone, even when applied to a distinct audio domain \cite{DASS}.

Next, we evaluate the impact of knowledge distillation. Without any distillation (1-index), the model only achieves 45.3\% mAP, while enabling knowledge distillation stages progressively improves performance. For instance, 2-index (IN + Stage 1) and 3-index (IN + Stages 1–2) achieve 47.2\% and 47.5\% mAP, respectively. The performance gain in Stage 2, where the SSM backbone is frozen and auxiliary branches are introduced, indicates that training local enhancement modules under CNN-guided distillation helps improve feature extraction and discriminative representation learning. Replacing the CNN teacher with a Transformer-based teacher in the final stage (Stage 3) yields notable improvements. The 4-index (IN + Stage 3) and 5-index (IN + Stages 2–3) achieve 49.9\% and 50.1\% mAP, respectively, outperforming the CNN-only distillation setting in 3-index (47.5\%). These results demonstrate the benefit of Transformer-guided distillation for global contextual modeling.

Finally, the full multi-stage distillation (6-index) attains the highest mAP (50.3\%). This confirms the effectiveness of the architecture-complementary design: sequentially combining CNN and Transformer teachers during distillation enables the Mamba-based model to preserve its computational efficiency while integrating both local and global modeling strengths, ultimately maximizing its discriminative capacity.

\begin{table}[t]
    \centering
    \caption{Ablation study on ImageNet initialization and multi-stage knowledge distillation in AD-Mamba(S). IN: ImageNet initialization}
    \footnotesize
    \setlength{\tabcolsep}{4.5pt}
    \label{ablation IN and KD}
    \begin{tabular}{c|cccc|c}
        \hline
        Index & IN & Stage1 & Stage2 & Stage3 & mAP \\
        \hline
        0 & False & True & True & True & 48.9 \\
        1 & True & False & False & False & 45.3 \\
        2 & True & True & False & False & 47.2 \\
        3 & True & True & True & False & 47.5 \\
        4 & True & False & False & True & 49.9 \\
        5 & True & False & True & True & 50.1 \\
        \hline
        6 & True & True & True & True & \textbf{50.3} \\
        \hline
    \end{tabular}
    \vspace{-0.1cm}
\end{table}

\textbf{Training Duration Analysis}: To validate that the performance gain is due to the architecture-complementary multi-stage design rather than simply an extended training duration, we conduct a controlled experiment summarized in Table~\ref{ablation_extended_epochs}. We extend the training of the Transformer-Only knowledge distillation approach to match the total epoch count of the complete multi-stage process. The extended training for the Transformer-Only model (1-index) yields only 49.7\% mAP, which is not only a marginal decrease from its standard 200-epoch result (49.9\% mAP) but also significantly lower than the result achieved by the full multi-stage approach (50.3\% mAP). This demonstrates that merely prolonging training duration is insufficient to achieve effective knowledge transfer, and suggests the importance of the CNN-guided stage in providing local inductive bias to the Mamba student model.

\subsubsection{Necessity of CNN distillation and multi-stage design}
To further validate the benefits of the CNN distillation stage, we compare the performance of the full AD-Mamba model, the CNN-Only distilled model, and the Transformer-Only distilled model on a noise robustness task. This task is designed to simulate performance maintenance when acoustic events are sparsely distributed amidst noise interference, with the specific setup consistent with that of Figure~\ref{figure6}. Detailed results are presented in Table~\ref{ablation robustness}.

While the Transformer-Only distilled model initially achieves a significantly higher baseline performance (49.9\% mAP at 0s noise padding, compared to 47.5\% mAP for the CNN-Only model, a difference of 2.4\%), the CNN-Only distilled model exhibits slower performance decay under prolonged noise interference. Although the CNN-Only model never fully surpasses the absolute mAP of the Transformer-Only model, the performance gap narrows substantially as noise duration increases. At 100 seconds of noise padding, this gap shrinks to a mere 0.4\% (32.3\% vs. 32.7\% mAP). These findings demonstrate that the CNN teacher successfully transfers essential local feature stability, providing better robustness to temporal interference compared to the purely global modeling approach of the Transformer teacher.

Furthermore, we observe that the Full Multi-Stage model maintains an absolute leading performance. While the Transformer-Only model's initial performance is close (only 0.4\% behind at 0s), the performance gap widens to 2.5\% at 100 seconds of noise padding (32.7\% vs. 35.2\% mAP). This supports the importance of integrating local inductive biases and global modeling capabilities for constructing models that achieve both high classification performance and strong robustness.

\begin{table}[t]
    \centering
    \caption{Impact of training duration. Comparison of Transformer-Only knowledge distillation with extended training duration against the full multi-stage approach.}
    \footnotesize
    \label{ablation_extended_epochs}
    \begin{tabular}{c|cc|c}
        \hline
        Index & Teacher Architecture & Total Epochs & mAP \\
        \hline
        0 & Transformer-Only & 200 & 49.9 \\
        1 & Transformer-Only & 400 (Extended) & 49.7 \\
        2 & CNN $\rightarrow$ Transformer & 400 & \textbf{50.3} \\
        \hline
    \end{tabular}
\end{table}

\begin{table}[t]
    \centering
    \caption{Robustness comparison of different distillation strategies under the noise padding task.}
    \label{ablation robustness}
    \footnotesize
    \setlength{\tabcolsep}{2pt}
    \begin{tabular}{c|cccc}
        \hline
        Distillation Strategy & mAP (0s) & mAP (10s) & mAP (50s) & mAP (100s) \\
        \hline
        CNN-Only & 47.5 & 39.7 & 35.6 & 32.3 \\
        Transformer-Only & 49.9 & 41.5 & 36.4 & 32.7 \\
        Full Multi-Stage & \textbf{50.3} & \textbf{42.8} & \textbf{38.8} & \textbf{35.2} \\
        \hline
    \end{tabular}
\end{table}

We next validate the necessity of our sequential CNN distillation design (Stage 1 and Stage 2) through a controlled experiment, with results presented in Table \ref{ablation stages}. In the comparative setup, we bypass phased optimization and directly train all modules of the AD-Mamba student model simultaneously under the guidance of the CNN teacher (Direct CNN Distillation). To ensure fairness, this stage involves 200 training epochs, matching the combined duration of Stage 1 and Stage 2. Following the subsequent Transformer distillation stage, the direct CNN distillation strategy achieves 50.0\% mAP, which is 0.3\% lower than that of the full multi-stage knowledge distillation. One possible explanation is that simultaneously optimizing all student parameters in the initial phase may dilute the learning focus, making it more difficult for local feature modules to effectively capture fine-grained local knowledge from the CNN teacher. In contrast, our sequential design ensures targeted optimization: Stage 1 adapts the core Mamba sequence component to the audio classification task, and Stage 2 introduces and fine-tunes local feature modules, maximizing their capacity to assimilate the CNN's local feature extraction capabilities.

\subsubsection{Importance of strong teacher and student models}

Table~\ref{strong teacher and student} presents an ablation study investigating the impact of teacher and student model capabilities on distillation performance. The left half of the table compares two teacher models: ASDA, a strong pre-trained model, and AudioMAE, a weaker baseline. The right half contrasts two student architectures: our proposed AD-Mamba and DASS, a representative Mamba-based audio model.

The results indicate that improving either the teacher or the student model yields performance gains. Specifically, when using the DASS student, switching from the weaker AudioMAE teacher (0-index) to the stronger ASDA teacher (1-index) yields a 1.7\% mAP improvement. With the teacher model fixed, replacing DASS with AD-Mamba as the student (2-index) increases mAP by 0.6\%. Similarly, under the stronger ASDA teacher, upgrading the student model from DASS to AD-Mamba (3-index) still results in a 0.5\% mAP gain.

Overall, the transition from the weakest configuration (0-index) to the strongest (3-index) achieves a total mAP improvement of 2.2\%, validating the importance of both a powerful teacher and a high-capacity student architecture in maximizing distillation performance for audio classification.

\begin{table}[t]
    \centering
    \caption{Ablation study on the efficacy of sequential CNN distillation stages.}
    \label{ablation stages}
    \footnotesize
    \setlength{\tabcolsep}{2pt}
    \begin{tabular}{c|cc|c}
        \hline
        Index & CNN Distillation & Transformer Distillation & mAP \\
        \hline
        0 & Direct CNN Distillation & Applied & 50.0 \\
        1 & Sequential (Stage 1 $\rightarrow$ Stage 2) & Applied  & 50.3 \\
        \hline
    \end{tabular}
\end{table}

\begin{table}[t]
    \centering
    \caption{Ablation study on strong teacher and student models}
    \label{strong teacher and student}
    \footnotesize
    \begin{tabular}{c|cc|cc|c}
        \hline
        & \multicolumn{2}{c|}{Teacher Model} & \multicolumn{2}{c|}{Student Model} & \\
        \cline{2-5}
        Index & ASDA & AudioMAE & AD-Mamba & DASS & mAP \\
        \hline
        0 & False & True & False & True & 47.7 \\
        1 & True & False & False & True & 49.4 \\
        2 & False & True & True & False & 48.3 \\
        3 & True & False & True & False & 49.9 \\
        \hline
    \end{tabular}
    \vspace{-0.1cm}
\end{table}

\subsubsection{Importance of model and parameter configuration}

To validate the impact of architectural components and hyperparameters in AD-Mamba(S), we conduct ablation experiments as detailed in Table \ref{ablation model and parameter}.

We first assess the importance of the spatial and channel (SC) branches and the local perception unit (LPU). As shown in 0-index and 1-index, individually removing either the SC or LPU leads to a performance drop of 0.2\% mAP compared to the full model (7-index). This suggests that both local feature extraction (via LPU) and cross-dimension feature interaction (via SC) play complementary roles in enhancing discriminative capacity. Enabling both yields the best performance (50.3\% mAP), suggesting that their integration improves representation learning for audio events.

Next, we evaluate the design of SSM from the perspective of scanning directionality. Specifically, we compare unidirectional, bidirectional, and quad-directional configurations while keeping the parameter count constant by replicating branches where necessary. As indicated by index 2–3 and 7, the quad-directional setup achieves the best performance (50.3\% mAP), whereas unidirectional and bidirectional variants slightly underperform. This result highlights the advantage of incorporating richer contextual flows from multiple directions, which helps the model perceive complex time-frequency patterns.

We also explore the effect of two critical knowledge distillation hyperparameters: the temperature $\tau$ and the loss balancing factor $\lambda$. As shown in index 4–7, setting $\tau = 2.0$ and $\lambda = 0.5$ leads to the best balance between ground truth supervision and soft target guidance. Deviating from this balance, either by reducing the temperature to $\tau = 1.0$ (4-index) or adjusting $\lambda$ to 0.3 or 0.7 (index 5-6), slightly impairs the model’s performance. These observations confirm that an appropriate distillation temperature helps smooth the logits of the teacher and emphasize informative gradients, while a moderate $\lambda$ encourages synergy between hard and soft targets.

\begin{table}[t]
    \centering
    \caption{Ablation study on model and parameter configuration in AD-Mamba(S).}
    \label{ablation model and parameter}
    \footnotesize
    \begin{tabular}{c|cccccc|c}
        \hline
        Index & SC & LPU & SSM &Params & $\tau$ & $\lambda$ & mAP \\
        \hline
        0 & \ding{55} & \ding{51} & Quad & 30M & 2.0 & 0.5 & 50.1 \\
        1 & \ding{51} & \ding{55} & Quad & 30M & 2.0 & 0.5 & 50.1 \\
        2 & \ding{51} & \ding{51} & Unidi & 30M & 2.0 & 0.5 & 49.8 \\
        3 & \ding{51} & \ding{51} & Bidi & 30M & 2.0 & 0.5 & 50.1 \\
        4 & \ding{51} & \ding{51} & Quad & 30M & 1.0 & 0.5 & 50.1 \\
        5 & \ding{51} & \ding{51} & Quad & 30M & 2.0 & 0.3 & 50.0 \\
        6 & \ding{51} & \ding{51} & Quad & 30M & 2.0 & 0.7 & 50.2 \\ 
        \hline
        7 & \ding{51} & \ding{51} & Quad & 30M & 2.0 & 0.5 & \textbf{50.3} \\
        \hline
    \end{tabular}
    \vspace{-0.1cm}
\end{table}

\section{Downstream task experiments and results}
\label{downstream}
To assess the generalization and transferability of the proposed method, we conduct fine-tuning experiments across three representative audio tasks: sound event tagging, environmental sound classification, and acoustic scene classification.

\subsection{Datasets}

\subsubsection{Sound event tagging on FSD50K}

The FSD50K \cite{FSD50K} dataset comprises 37,134 training clips, 4,170 validation clips, and 10,231 testing audio clips, with durations ranging from 0.3 to 30 seconds and totaling approximately 100 hours. With 200 event categories, each audio clip can be associated with one or more labels, thus we adopt mean average precision (mAP) as the evaluation metric to account for the multi-label nature of the task.

\subsubsection{Environmental sound classification on ESC-50}

The ESC-50 \cite{ESC-50} dataset includes 2,000 balanced 5-second audio clips across 50 environmental sound classes. Each clip is associated with a single label. Following official protocols, we perform 5-fold cross-validation, where each fold comprises 1,600 training clips and 400 testing clips, with results reported as the average classification accuracy across all folds. 

\subsubsection{Acoustic scene classification on DCASE16}

The TUT Acoustic Scenes 2016 dataset (DCASE16) \cite{DCASE16} contains 1,170 training clips and 390 testing clips, each 30 seconds in duration, totaling around 13 hours. The dataset is composed of 15 acoustic scenes, with balanced sampling across training and testing stages, with each scene including 78 training and 26 testing clips. Accuracy serves as the evaluation metric for this single-label classification task.

\subsection{Fine-Tuning experimental setup}

The downstream fine-tuning experiments maintain identical data preprocessing and augmentation strategies as described in Section~\ref{data processing}, with task-specific adjustments only to training epochs, learning rates, and schedulers. 

For the FSD50K task, we train the model for 30 epochs: exponential warm-up (5 epochs), constant learning rate phase (15 epochs at 1e-4), and linear decay (10 epochs, decaying by 0.8 every 2 epochs). For ESC-50, training is conducted for 100 epochs: warm-up (5 epochs), constant rate (15 epochs at 1e-4), and decay (80 epochs, 0.8 decay every 4 epochs). For DCASE16, the scheduler includes warm-up (15 epochs), constant rate (35 epochs at 5e-5), and decay (50 epochs, decaying by 0.8 every 4 epochs).

All downstream experiments utilize the Adam optimizer, with a weight decay coefficient of 0 and a batch size of 48.

\begin{table}[t]
    \centering
    \caption{Performance comparison with representative baseline models on the downstream tasks FSD50K, ESC-50 and DCASE16.}
    \label{downstream comparison}
    \footnotesize
    \setlength{\tabcolsep}{2pt}
    \begin{tabular}{c|cccc}
        \hline
        Model & Params & FSD50K & ESC-50 & DCASE16 \\
        \hline 
        MN\cite{MN-KD} & 68M & 63.1 & 96.2 & 81.3 \\
        DyMN (16\,kHz)\cite{kd-cnn} & 40M & 62.3 & 95.1 & 90.2 \\
        PaSST\cite{passt} & 86M & 65.3 & 96.8 & 78.8 \\
        EAT\cite{EAT} & 88M & 65.1 & 95.9 & 90.8 \\
        ASDA\cite{ASDA} & 93M & 65.4 & 96.1 & 90.9 \\
        CED\cite{CED} & 86M & 65.5 & 96.7 & 92.2 \\
        \hline
        AD-Mamba(S) & 30M & \textbf{68.7} & \textbf{96.9} & \textbf{93.3} \\ 
        \hline
    \end{tabular}
    \vspace{-0.1cm}
\end{table}

\subsection{Comparison of fine-tuning results}

Table \ref{downstream comparison} presents the performance comparison of our proposed AD-Mamba(S) against representative baselines across three downstream tasks. Among them, MN \cite{MN-KD}, PaSST \cite{passt} and CED \cite{CED} are trained using 32\,kHz, whereas all other methods, including ours, are trained at 16\,kHz. Although higher sampling rates can provide richer high-frequency information, they also incur greater computational and memory costs.

On the FSD50K sound event tagging task, AD-Mamba(S) achieves 68.7\% mAP, outperforming all prior models, such as ASDA (65.4\% at 16\,kHz), and CED (65.5\% at 32\,kHz). This performance gain suggests strong multi-label event tagging capability, enabling accurate identification of multiple concurrent sound events within complex audio clips.

For the ESC-50 environmental sound classification task, AD-Mamba(S) obtains an accuracy of 96.9\%, marginally surpassing 32\,kHz models PaSST (96.8\%) and CED (96.7\%), and significantly outperforming the previous best 16\,kHz baseline ASDA (96.1\%). These results highlight the strong transferability of the model, even in tasks requiring high-frequency resolution to distinguish subtle environmental sound differences.

On the DCASE16 acoustic scene classification task, AD-Mamba(S) achieves 93.3\% accuracy, obtaining the best performance among the compared methods and exceeding CED (92.2\%), despite using fewer parameters (30M vs. 86M) and a lower sampling rate. This result demonstrates the effectiveness and efficiency of the proposed model for complex acoustic scene perception.

In summary, AD-Mamba(S) achieves the best performance among the compared models across all three downstream tasks while operating at 16\,kHz sampling rate. Coupled with its significantly reduced parameter count, the model offers a favorable balance between computational efficiency and classification performance, supporting its potential for practical audio classification scenarios.

\section{Conclusion}
\label{conclusion}

In this work, we introduce AD-Mamba, an efficient Mamba-based audio classification model trained with architecture-\allowbreak{}com\-ple\-men\-ta\-ry multi-stage knowledge distillation. By leveraging existing CNN and Transformer models pre-trained on large-scale datasets as teachers, our framework sequentially transfers the local perception and translation equivariance characteristics of CNNs, as well as the global contextual modeling strengths of Transformers, into the AD-Mamba framework. When combined with the inherent linear-time sequence modeling capability and efficient parallel processing of Mamba, this integration enables AD-Mamba to achieve strong accuracy-efficiency trade-offs across various audio tasks.

Specifically, AD-Mamba achieves 50.3\% mAP on AudioSet with only 30M parameters, and the performance further improves to 50.6\% mAP when scaled to 49M parameters, outperforming representative CNN-, Transformer-, and Mamba-based audio classification baselines. Notably, compared to CNN and Transformer baselines, this method demonstrates stronger capabilities in handling long-form audio, extracting sparse audio information, and maintaining robustness against noise interference. When transferred to downstream tasks, AD-Mamba exhibits strong generalization capability across different audio perception domains. Even when operating at a lower sampling rate of 16\,kHz, it achieves competitive or superior performance compared with models trained with 32\,kHz inputs. This result highlights its ability to deliver exceptional performance under reduced computational costs and parameter budgets.

Overall, AD-Mamba provides evidence that SSM-based architectures can serve as competitive alternatives to attention-based models for large-scale audio classification when guided by appropriate knowledge distillation strategies. Different from previous audio Mamba models that mainly explored the feasibility and efficiency of SSMs in audio tasks, this work demonstrates that architecture-complementary distillation can substantially enhance the discriminative capability of Mamba-based audio classifiers while preserving their computational advantages. Beyond performance improvements, the compact design and robustness of AD-Mamba further indicate its potential for resource-constrained audio classification scenarios, particularly when long or noisy audio streams need to be processed efficiently.

\bibliographystyle{IEEEtran}
\bibliography{references}

\end{document}